# Reciprocity in Social Networks with Capacity Constraints

Bo Jiang
College of Information and
Computer Sciences
University of Massachusetts
Amherst MA, USA
bjiang@cs.umass.edu

Zhi-Li Zhang
Department of Computer
Science and Engineering
University of Minnesota
Minneapolis MN, USA
zhzhang@cs.umn.edu

Don Towsley
College of Information and
Computer Sciences
University of Massachusetts
Amherst MA, USA
towsley@cs.umass.edu

## ABSTRACT

*Directed* links – representing *asymmetric* social ties or interactions (e.g., "follower-followee") – arise naturally in many social networks and other complex networks, giving rise to directed graphs (or digraphs) as basic topological models for these networks. *Reciprocity*, defined for a digraph as the percentage of edges with a *reciprocal* edge, is a key metric that has been used in the literature to compare different directed networks and provide "hints" about their structural properties: for example, are reciprocal edges generated randomly by chance or are there other processes driving their generation? In this paper we study the problem of *maximizing achievable reciprocity* for an ensemble of digraphs with the same prescribed in- and out-degree sequences. We show that the maximum reciprocity hinges crucially on the in- and out-degree sequences, which may be intuitively interpreted as constraints on some "social capacities" of nodes and impose fundamental limits on achievable reciprocity. We show that it is NP-complete to decide the achievability of a simple upper bound on maximum reciprocity, and provide conditions for achieving it. We demonstrate that many real networks exhibit reciprocities surprisingly close to the upper bound, which implies that users in these social networks are in a sense more "social" than suggested by the empirical reciprocity alone in that they are more willing to reciprocate, subject to their "social capacity" constraints. We find some surprising linear relationships between empirical reciprocity and the bound. We also show that a particular type of small network motifs that we call 3-paths are the major source of loss in reciprocity for real networks.

## Categories and Subject Descriptors

G.2.2 [**Graph Theory**]: Network problems; H.2.8 [**Database Applications**]: Data mining

## 1. INTRODUCTION

Many complex networks are naturally directed, which endows them with nontrivial structural properties not shared by undirected networks. One such property that has been widely studied is *reciprocity*, which is classically defined as the fraction of edges that are reciprocated, i.e. paired with an edge of the opposite direction. Nontrivial patterns of reciprocity can reveal possible mechanisms of social, biological or other nature that systematically act as organizing principles shaping the observed network topology [6]. Previous work shows that reciprocity plays an important role in many information networks such as email networks [16], the World Wide Web [2] and Wikipedia [29, 28]. It is also shown that major online social networks that are directed in nature, such as Twitter[10, 12], Google+[14], Flickr [15, 4], LiveJournal [25, 15, 7], and YouTube [15], all exhibit a nontrivial amount of reciprocity.

When we try to interpret observed values of reciprocity, we are faced with the problem of assessing the significance of the observation. For instance, the Swedish Wikipedia has reciprocity of 21%. How significant is this? This question is often answered by comparing measured values with the expected value of some null model. One commonly used null model is a random graph with the same number of nodes and edges [16]. An alternative is a random graph with specified degree sequence, as the specific degree sequence is expected to affect reciprocity [26]. Networks are then classified as *reciprocal* or *anti-reciprocal* according to whether the observed reciprocity is larger or smaller than the expected value [6]. Significant deviation from the expected values suggests the existence of some underlying organizational mechanism at work. For our example of Swedish Wikipedia, the expected reciprocities in both random null models are almost zero, so the Swedish Wikipedia is classified as a reciprocal network. Informative as this might be, comparison with expected values is not the whole story. Is 21% a significant deviation from 0? Can we say that the tendency to reciprocate is strong in this network? The answer might depend on the eye of the beholder. However, if we know for some reason the maximum possible reciprocity is only 28%, then we may safely conclude that 21% is indeed a significant amount of reciprocity. On the other hand, if the maximum is 90%, we might conclude that 21% is not as significant as suggested by the comparison with random null models. In general, knowledge of the extremal values can give a better idea about where the observation lies in the entire spectrum, which can potentially change our conclusion about the significance level of the observation.

Since real social networks often exhibit reciprocities larger than those associated with the random null models, we concern ourselves only with the maximum achievable reciprocity in this work. As in the random null models, we may want to retain certain key structural features of the real network when we maximize reciprocity. The particular feature that we choose to preserve in this work is the joint in- and out-degree sequence, which is a confounding factor in the study of reciprocity [26]. In real networks, in- and out-degrees often serve as proxies for some kind of capacities of the corresponding node. For example, in a file sharing network where edges represent transfers from file sources to downloaders, the in-degree of a node can reflect the available network bandwidth and the out-degree the amount of resource. In a social network where edges point from followers to followees, the in-degree of a node can reflect its fame and popularity and the out-degree its budget of attention. Quite often these capacity constraints are too important to be ignored in the network under consideration. By preserving the degree sequence, we honor these capacity constraints, thus controlling these confounding factors.

Motivated by the above considerations, we study the problem of maximizing reciprocity subject to prescribed joint in- and out-degree constraints. This paper makes the following contributions.

- We formulate the *maximum reciprocity problem* and prove its NP-hardness. We provide an upper bound on reciprocity and conditions for achieving the bound.

- We show that empirical reciprocity is surprisingly close to the upper bound in a wide range of real networks. We also find surprisingly strong linear relationships between empirical reciprocity and the upper bound.

- We identify some suboptimal network motifs and show that a particular type of small motif called 3-paths is the major cause for suboptimality in real networks.

The rest of the paper is organized as follows. Section 2 introduces the *maximum reciprocity problem*. Section 3 proves the NP-hardness of the problem, and provides a simple upper bound for maximum reciprocity. Section 4 identifies patterns of maximum digraphs and provides a greedy algorithm for eliminating suboptimal motifs. Section 5 conducts some empirical study of real networks and Section 6 concludes the paper.

## 2. GRAPHIC SEQUENCES AND MAXIMUM RECIPROCITY PROBLEM

In this section, we first introduce the notion of a *graphic sequence* for undirected graphs and then a *graphic bi-sequence* for directed graphs or digraphs for short, which will be used in the theoretical analysis of Section 3. We then formulate the maximum reciprocity problem. Throughout the rest of the paper, a graph, directed or not, always means a simple graph, i.e. no self-loops or multiple edges are allowed. We will use the terms *node* and *vertex* interchangeably. For directed graphs, an *edge* always means a *directed edge*.

### 2.1 Graphic Sequence and Bi-sequence

For an undirected graph $G = (V, E)$, the degree $d_G(v)$ of a node $v$ is the number of edges incident to $v$. Associated with every graph $G$ is its degree sequence $\mathbf{d} = \{d_G(v) : v \in V\}$. However, not every sequence of nonnegative integers can be realized as the degree sequence of a graph. When it is realizable, the sequence is called *graphic*. More precisely, a sequence of nonnegative integers $\mathbf{d} = (d_1, d_2, \ldots, d_n)$ is called *graphic* if there exists a graph $G$ with nodes $v_1, v_2, \ldots, v_n$ such that $d_G(v_i) = d_i$ for $i = 1, 2, \ldots, n$. The following classical theorem of Erdős and Gallai characterizes graphic sequences.

THEOREM 1 (ERDŐS-GALLAI). *A sequence of nonnegative integers $d_1 \geq d_2 \geq \cdots \geq d_n$ is graphic if and only if $\sum_{i=1}^{n} d_i$ is even and*

$$\sum_{i=1}^{k} d_i \leq k(k-1) + \sum_{i=k+1}^{n} \min\{d_i, k\}, \quad \text{for } k = 1, 2, \ldots, n.$$

The graphicality of a sequence can be tested in linear time using the Erdős-Gallai theorem [9].

For a digraph $G = (V, E)$, a node has both an in-degree and an out-degree. The in-degree $d_G^-(v)$ of a node $v$ is the number of directed edges coming into $v$, and the out-degree $d_G^+(v)$ is the number of directed edges going out of $v$. Associated with every digraph $G$ is a *bi-sequence* $(\mathbf{d}^+, \mathbf{d}^-)$, where $\mathbf{d}^+ = \{d_G^+(v) : v \in V\}$ is the out-degree sequence and $\mathbf{d}^- = \{d_G^-(v) : v \in V\}$ is the in-degree sequence. As in the undirected case, not every bi-sequence of nonnegative integers can be realized by a digraph. A bi-sequence of nonnegative integers $(\mathbf{d}^+, \mathbf{d}^-) = \{(d_1^+, d_2^+, \ldots, d_n^+), (d_1^-, d_2^-, \ldots, d_n^-)\}$ is called *graphic* if there exists a digraph $G$ with nodes $v_1, v_2, \ldots, v_n$ such that $d_G^+(v_i) = d_i^+$ and $d_G^-(v_i) = d_i^-$ for $i = 1, 2, \ldots, n$. The Fulkerson-Chen-Anstee theorem characterizes graphic bi-sequences.

THEOREM 2 (FULKERSON-CHEN-ANSTEE). *A bi-sequence $\{(d_1^+, \ldots, d_n^+), (d_1^-, \ldots, d_n^-)\}$ with $d_1^+ \geq d_2^+ \geq \cdots \geq d_n^+$ is graphic if and only if $\sum_{i=1}^{n} d_i^+ = \sum_{i=1}^{n} d_i^-$ and*

$$\sum_{i=1}^{k} d_i^+ \leq \sum_{i=1}^{k} \min\{d_i^-, k-1\} + \sum_{i=k+1}^{n} \min\{d_i^-, k\},$$

*for $k = 1, 2, \ldots, n$.*

### 2.2 Maximum Reciprocity Problem

In this subsection, we formulate the maximum reciprocity problem. For notational simplicity, we henceforth make no distinction between a graph (digraph) and its edge set when no confusion arises.

Given a digraph $G$, let $G_s$ be the *symmetric* subgraph of $G$, i.e. $(i, j) \in G_s$ if and only if both $(i, j) \in G$ and $(j, i) \in G$. The reciprocated edges of a digraph $G$ are precisely those of $G_s$. Thus the number $\rho(G)$ of reciprocated edges in $G$ is given by $\rho(G) = |G_s|$, and the *reciprocity* of $G$ is $r(G) := \rho(G)/|G|$. Note that we use $|G|$ to denote the number of edges in $G$ and each pair of reciprocal edges contributes two to $\rho(G)$.

Given a graphic bi-sequence $(\mathbf{d}^+, \mathbf{d}^-)$, let $\mathcal{G}(\mathbf{d}^+, \mathbf{d}^-)$ denote the nonempty set of graphs that have $(\mathbf{d}^+, \mathbf{d}^-)$ as their degree bi-sequence. Since the total number of edges is fixed for a given graphic bi-sequence, maximizing $r(G)$ is the same as maximizing $\rho(G)$. The *maximum reciprocity problem* is then to find a digraph $G$ in $\mathcal{G}(\mathbf{d}^+, \mathbf{d}^-)$ with maximum $\rho(G)$, i.e.

$$\text{maximize } \rho(G)$$
$$\text{subject to } G \in \mathcal{G}(\mathbf{d}^+, \mathbf{d}^-).$$

We denote the maximum value by $\rho(\mathbf{d}^+, \mathbf{d}^-)$ and refer to a digraph $G$ with $\rho(G) = \rho(\mathbf{d}^+, \mathbf{d}^-)$ as a *maximum reciprocity digraph* or *maximum digraph* for short.

## 2.3 Some Notations

We collect here some notations for later reference. Let $G$ denote a generic digraph.

- Let $G_a$ be the anti-symmetric subgraph of $G$, i.e. $(i,j) \in G_a$ if and only if $(i,j) \in G$ but $(j,i) \notin G$. Note that $G = G_s + G_a$ and $G_s \cap G_a = \emptyset$, i.e. $G$ is the edge disjoint union of $G_s$ and $G_a$.

- Let $G_u$ be the undirected graph obtained by symmetrizing $G$, i.e. $(i,j) \in G_u$ if either $(i,j) \in G$ or $(j,i) \in G$.

Let $(\mathbf{d}^+, \mathbf{d}^-)$ be a graphic bi-sequence.

- The min sequence is
$$\mathbf{d}^+ \wedge \mathbf{d}^- = (d_1^+ \wedge d_1^-, d_2^+ \wedge d_2^-, \ldots, d_n^+ \wedge d_n^-),$$
where $a \wedge b = \min\{a, b\}$.

- The max sequence is
$$\mathbf{d}^+ \vee \mathbf{d}^- = (d_1^+ \vee d_1^-, d_2^+ \vee d_2^-, \ldots, d_n^+ \vee d_n^-),$$
where $a \vee b = \max\{a, b\}$.

- The total number of edges is
$$\varepsilon(\mathbf{d}^+, \mathbf{d}^-) = \sum_i d_i^+ = \sum_i d_i^-.$$

- The total balanced degree is
$$\beta(\mathbf{d}^+, \mathbf{d}^-) = \sum_i d_i^+ \wedge d_i^-,$$
which is the $\ell_1$-norm of the min sequence.

- The total unbalanced degree is
$$\nu(\mathbf{d}^+, \mathbf{d}^-) = \frac{1}{2} \sum_i |d_i^+ - d_i^-|,$$
which is the total variation distance between $\mathbf{d}^+$ and $\mathbf{d}^-$. Note that $\varepsilon(\mathbf{d}^+, \mathbf{d}^-) = \beta(\mathbf{d}^+, \mathbf{d}^-) + \nu(\mathbf{d}^+, \mathbf{d}^-)$.

## 3. HARDNESS ANALYSIS AND BOUNDS

In this section, we first provide an upper bound for the maximum number of reciprocated edges allowed by a graphic bi-sequence. We then prove that the maximum reciprocity problem is NP-hard by showing that it is NP-complete to decide the achievability of the upper bound. Some sufficient conditions for achieving the upper bound are then provided.

### 3.1 Upper Bound for Reciprocity

In this subsection, we first establish a simple upper bound on the maximum number of reciprocal edges in terms of the total imbalance of the graphic bi-sequence, along with necessary conditions for achieving this upper bound. Some examples are provided to illustrate how the necessary conditions may fail and that they are not sufficient, which provides insight into why the bound is not always tight.

PROPOSITION 1. *The number of reciprocated edges in any digraph with a given degree bi-sequence cannot exceed the total balanced degree, i.e.*
$$\rho(\mathbf{d}^+, \mathbf{d}^-) \leq \beta(\mathbf{d}^+, \mathbf{d}^-).$$
*A necessary condition for equality is that both $\mathbf{d}^+ \wedge \mathbf{d}^-$ and $\mathbf{d}^+ \vee \mathbf{d}^-$ be graphic.*

PROOF. Let $G \in \mathcal{G}(\mathbf{d}^+, \mathbf{d}^-)$ be a maximum digraph. Note that the number of reciprocated edges going out of a node $v$ is at most $d_G^+(v) \wedge d_G^-(v)$. The desired bound is obtained by summing over $v$.

If equality holds, then $G_s$ and $G_u$, when viewed as undirected graphs, have respective degree sequences $\mathbf{d}^+ \wedge \mathbf{d}^-$ and $\mathbf{d}^+ \vee \mathbf{d}^-$. Thus both $\mathbf{d}^+ \wedge \mathbf{d}^-$ and $\mathbf{d}^+ \vee \mathbf{d}^-$ are graphic. □

Note that it is possible that neither $\mathbf{d}^+ \wedge \mathbf{d}^-$ nor $\mathbf{d}^+ \vee \mathbf{d}^-$ is graphic. In fact, one sequence can fail to be graphic independent of whether the other is graphic or not, as illustrated by the following examples, where graphic bi-sequences are shown along with the corresponding maximum digraphs.

*Example 1.* In Figure 1, neither the min sequence $\mathbf{d}^+ \wedge \mathbf{d}^-$ nor the max sequence $\mathbf{d}^+ \vee \mathbf{d}^-$ is graphic, since they both have odd sums. Here $\rho(\mathbf{d}^+, \mathbf{d}^-) = 2 < \beta(\mathbf{d}^+, \mathbf{d}^-) = 3$.

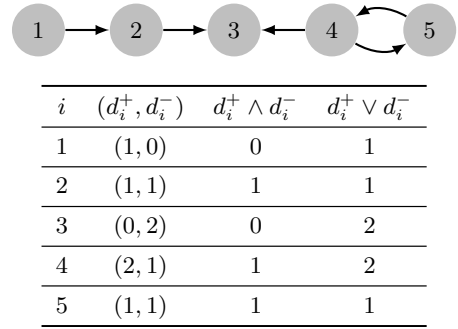

| $i$ | $(d_i^+, d_i^-)$ | $d_i^+ \wedge d_i^-$ | $d_i^+ \vee d_i^-$ |
|---|---|---|---|
| 1 | $(1, 0)$ | 0 | 1 |
| 2 | $(1, 1)$ | 1 | 1 |
| 3 | $(0, 2)$ | 0 | 2 |
| 4 | $(2, 1)$ | 1 | 2 |
| 5 | $(1, 1)$ | 1 | 1 |

Figure 1: Graphic bi-sequence with non-graphic max and min sequences.

*Example 2.* In Figure 2, the min sequence $\mathbf{d}^+ \wedge \mathbf{d}^-$ is graphic, while the max sequence $\mathbf{d}^+ \vee \mathbf{d}^-$ is not. No reciprocity is allowed by this bi-sequence, i.e. $\rho(\mathbf{d}^+, \mathbf{d}^-) = 0$, while the upper bound gives $\beta(\mathbf{d}^+, \mathbf{d}^-) = 2n$, so the gap can be arbitrarily large. The only unbalanced nodes $s$ and $r$ have very large unbalanced degrees that cannot be absorbed by themselves, as a consequence of which some, in fact all, balanced degrees have to be used for absorbing unbalanced degrees rather than forming reciprocal edges.

*Example 3.* In Figure 3, the max sequence $\mathbf{d}^+ \vee \mathbf{d}^-$ is graphic, while the min sequence $\mathbf{d}^+ \wedge \mathbf{d}^-$ is not. As in Example 2, no reciprocity is allowed here, i.e. $\rho(\mathbf{d}^+, \mathbf{d}^-)$, while the upper bound is $\beta(\mathbf{d}^+, \mathbf{d}^-) = 2n$. The situation is, however, the opposite. Node 0 has too large a balanced degree relative to the number of nodes with nonzero balanced degrees, which is one here. Thus some of the balanced degrees have to be absorbed by the unbalanced degrees.

The common pattern in Examples 2 and 3 is that there are a small number of nodes with extremely large degrees. In the social network context, these nodes correspond to celebrities (node $r$ in Figure 2), information aggregators (node $s$ in Figure 2), or middlemen (node 0 in Figure 3). These large degree nodes often incur inevitable reduction of reciprocity from the upper bound.

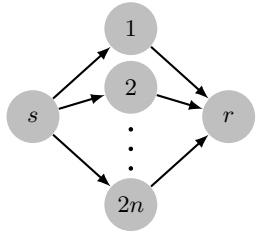

| $i$ | $(d_i^+, d_i^-)$ | $d_i^+ \wedge d_i^-$ | $d_i^+ \vee d_i^-$ |
| --- | --- | --- | --- |
| $s$ | $(2n, 0)$ | $0$ | $2n$ |
| $1 \sim 2n$ | $(1, 1)$ | $1$ | $1$ |
| $r$ | $(0, 2n)$ | $0$ | $2n$ |

Figure 2: Graphic bi-sequence with graphic min sequence but non-graphic max sequence.

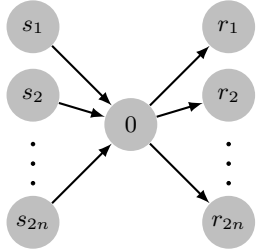

| $i$ | $(d_i^+, d_i^-)$ | $d_i^+ \wedge d_i^-$ | $d_i^+ \vee d_i^-$ |
| --- | --- | --- | --- |
| $s_1 \sim s_{2n}$ | $(1, 0)$ | $0$ | $1$ |
| $r_1 \sim r_{2n}$ | $(0, 1)$ | $0$ | $1$ |
| $0$ | $(2n, 2n)$ | $2n$ | $2n$ |

Figure 3: Graphic bi-sequence with graphic max sequence but non-graphic min sequence.

The next example shows that the necessary condition in Proposition 1 is not sufficient.

*Example 4.* For the bi-sequence $(d_i^+, d_i^+) = (n-i, i)$, $i = 0, 1, \ldots, n$, the upper bound is $\beta(\mathbf{d}^+, \mathbf{d}^-) = \lfloor n/2 \rfloor \cdot \lceil n/2 \rceil$. When $n$ is a multiple of 4, both the max sequence $\mathbf{d}^+ \vee \mathbf{d}^-$ and the min sequence $\mathbf{d}^+ \wedge \mathbf{d}^-$ are graphic. However, $\rho(\mathbf{d}^+, \mathbf{d}^-) = 0$, as the only digraph in $\mathcal{G}(\mathbf{d}^+, \mathbf{d}^-)$, of which $(i, j)$ is an edge if and only if $i < j$, has zero reciprocity; see Figure 4.

### 3.2 Proof of NP-hardness

We saw in the previous subsection that the upper bound may not be achievable. Unfortunately, the next theorem shows that it is NP-complete to decide whether the upper bound is achievable, which means the maximum reciprocity problem is NP-hard.

THEOREM 3. *The decision problem whether $\rho(\mathbf{d}^+, \mathbf{d}^-) = \beta(\mathbf{d}^+, \mathbf{d}^-)$ is NP-complete.*

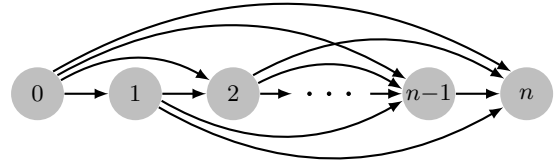

| $i$ | $(d_i^+, d_i^-)$ | $d_i^+ \wedge d_i^-$ | $d_i^+ \vee d_i^-$ |
| --- | --- | --- | --- |
| $0 \sim \lfloor n/2 \rfloor$ | $(n-i, i)$ | $i$ | $n-i$ |
| $\lceil n/2 \rceil \sim n$ | $(n-i, i)$ | $n-i$ | $i$ |

Figure 4: Insufficiency of necessary condition.

PROOF. Note that the problem is the same as the existence problem of a digraph $G \in \mathcal{G}(\mathbf{d}^+, \mathbf{d}^-)$ with $\rho(G) = \beta(\mathbf{d}^+, \mathbf{d}^-)$. This problem is in NP, since given a digraph $G$, we can verify whether $\rho(G) = \beta(\mathbf{d}^+, \mathbf{d}^-)$ in polynomial time. To show that the problem is NP-hard, we adapt the proof of Lemma 5 in [5] by reduction from the 3-color tomography problem, which is shown to be NP-hard therein.

Recall that the 3-color tomography problem is as follows. Given nonnegative integral vectors $r^w, r^b \in \mathbb{N}^n$, and $s^w, s^b \in \mathbb{N}^m$ that satisfy

$$r_i^w + r_i^b \leq m, \quad s_j^w + s_j^b \leq n, \quad \text{for } 1 \leq i \leq n, 1 \leq j \leq m,$$

and

$$\sum_{i=1}^n r_i^c = \sum_{j=1}^m s_j^c, \quad \text{for } c \in \{w, b\},$$

decide if $(r^w, r^b, s^w, s^b)$ is feasible, i.e. there exists a matrix $M$ with entries in $\{w, b, g\}$ such that

$$r_i^c = |\{j : M_{ij} = c\}|, \quad s_j^c = |\{i : M_{ij} = c\}|, \quad \text{for } c \in \{w, b\}.$$

Let $(r^w, r^b, s^w, s^b)$ be an $n \times m$ instance of the 3-color tomography problem. For $1 \leq i \leq n$ and $1 \leq j \leq m$, let

$$\begin{aligned} d_i^+ &= r_i^w + r_i^b + n - 1, & d_{n+j}^+ &= s_j^w, \\ d_i^- &= r_i^w + n - 1, & d_{n+j}^- &= s_j^w + s_j^b. \end{aligned} \quad (1)$$

Now we show that $(r^w, r^b, s^w, s^b)$ is feasible if and only if $(\mathbf{d}^+, \mathbf{d}^-)$ is graphic and $\rho(\mathbf{d}^+, \mathbf{d}^-) = \beta(\mathbf{d}^+, \mathbf{d}^-)$, where $\beta(\mathbf{d}^+, \mathbf{d}^-) = n(n-1) + 2\sum_{i=1}^n r_i^w$.

First assume that $M$ is a solution to the 3-color tomography instance. We construct a digraph $G$ as follows. For $1 \leq i \leq n$ and $1 \leq j \leq m$, let $W_{ij} = 1$ if $M_{ij} = w$, and $B_{ij} = 1$ if $M_{ij} = b$. Let $J$ be an $n \times n$ matrix with all off-diagonal entries equal to 1 and diagonal entries equal to 0. Let the adjacency matrix of $G$ be

$$\begin{pmatrix} J & W + B \\ W^T & 0 \end{pmatrix}.$$

It is straightforward to verify that $G \in \mathcal{G}(\mathbf{d}^+, \mathbf{d}^-)$ and $\rho(G) = \beta(\mathbf{d}^+, \mathbf{d}^-)$.

For the reverse direction, assume that $(\mathbf{d}^+, \mathbf{d}^-)$ is graphic and $\rho(\mathbf{d}^+, \mathbf{d}^-) = \beta(\mathbf{d}^+, \mathbf{d}^-)$. Then there exists a digraph $G \in \rho(\mathbf{d}^+, \mathbf{d}^-)$ with $\rho(G) = \beta(\mathbf{d}^+, \mathbf{d}^-)$. Divide the adjacency matrix of $G$ into the following block form

$$G = \begin{pmatrix} G_{11} & G_{12} \\ G_{21} & G_{22} \end{pmatrix}.$$

where $G_{11}$ is $n \times n$ and $G_{22}$ is $m \times m$.

Let $\Phi = \sum_{j=1}^n d_j^- - \sum_{i=1}^m d_{n+i}^+$, which, by (1), evaluates to $n(n-1)$. On the other hand, $d_j^- = \sum_{k=1}^{n+m} G(k,j)$ and $d_{n+i}^+ = \sum_{k=1}^{n+m} G(n+i,k)$, so

$$\Phi = \sum_{i=1}^n \sum_{j=1}^n G_{11}(i,j) - \sum_{i=1}^m \sum_{j=1}^m G_{22}(i,j) \le n(n-1) = \Phi,$$

where the inequality follows from the facts that $G_{11}(i,j) \le 1$, $G_{11}(i,i) = 0$ and $G_{22}(i,j) \ge 0$. Since the equality holds, we must have $G_{11} = J$ and $G_{22} = 0$. Thus

$$\rho(G) = n(n-1) + 2\sum_{i=1}^n \sum_{j=1}^m G_{12}(i,j) G_{21}(j,i)$$

$$\le n(n-1) + 2\sum_{i=1}^n \sum_{j=1}^m G_{21}(j,i)$$

$$= n(n-1) + 2\sum_{j=1}^m d_{n+j}^+ = \beta(\mathbf{d}^+, \mathbf{d}^-) = \rho(G).$$

Since the equality holds, $G_{12}(i,j) \ge G_{21}(j,i)$. Thus $G_{12} = W + B$ and $G_{21} = W^T$ for some $(0,1)$-matrices $W$ and $B$. Let $M_{ij} = w$ if $W(i,j) = 1$, and $M_{ij} = b$ if $B_{ij} = 1$. Then $M$ is a solution to the 3-color tomography instance.

Since the graphicality of $(\mathbf{d}^+, \mathbf{d}^-)$ can be tested in quadratic time using the Fulkerson-Chen-Anstee theorem, the above reduction then shows that it is NP-hard to decide whether $\rho(\mathbf{d}^+, \mathbf{d}^-) = \beta(\mathbf{d}^+, \mathbf{d}^-)$. □

### 3.3 Sufficient Conditions for Achieving Bound

Given the hardness of the maximum reciprocity problem, we provide some sufficient conditions for achieving the upper bound in Proposition 1. We start with the following slightly more general theorem, which may potentially be used to lower bound $\rho(\mathbf{d}^+, \mathbf{d}^-)$.

THEOREM 4. *Suppose that $\mathbf{d}^0$ is a graphic sequence such that the residual bi-sequence $(\mathbf{d}^+ - \mathbf{d}^0, \mathbf{d}^- - \mathbf{d}^0)$ is also graphic. If $\Delta < \sqrt{\delta n + \left(\delta - \frac{1}{2}\right)^2} + \frac{3}{2} - \delta$, where $n = |V_0|$, $\Delta = \bigvee_{i \in V_0}(d_i^+ + d_i^- - d_i^0)$ and $\delta = \bigwedge_{i \in V_0}(d_i^+ + d_i^- - d_i^0)$, with $V_0 = \{i : d_i^+ \vee d_i^- > 0\}$, then $\rho(\mathbf{d}^+, \mathbf{d}^-) \ge 2\sum_i d_i^0$.*

This theorem is analogous to Theorem 2.2 in [3], which deals with packing two graphic sequences for undirected graphs. Theorem 4 deals with packing a graphic sequence $\mathbf{d}^0$ for undirected graphs and a graphic bi-sequence $(\mathbf{d}^+ - \mathbf{d}^0, \mathbf{d}^- - \mathbf{d}^0)$ for digraphs. The proof closely follows that of Theorem 2.2 in [3] and can be found in [11].

Applying Theorem 4 with $\mathbf{d}^0 = \mathbf{d}^+ \wedge \mathbf{d}^-$, we obtain the following sufficient conditions for achieving the upper bound in Proposition 1.

COROLLARY 1. $\rho(\mathbf{d}^+, \mathbf{d}^-) = \beta(\mathbf{d}^+, \mathbf{d}^-)$ *if the following conditions hold,*

(1). $\mathbf{d}^+ \wedge \mathbf{d}^-$ and $(\mathbf{d}^+ - \mathbf{d}^+ \wedge \mathbf{d}^-, \mathbf{d}^- - \mathbf{d}^+ \wedge \mathbf{d}^-)$ are graphic;

(2). $\Delta < \sqrt{\delta n + \left(\delta - \frac{1}{2}\right)^2} + \frac{3}{2} - \delta$, where $n = |V_0|$, $\Delta = \bigvee_{i \in V_0}(d_i^+ \vee d_i^-)$ and $\delta = \bigwedge_{i \in V_0}(d_i^+ \vee d_i^-)$, with $V_0 = \{i : d_i^+ \vee d_i^- > 0\}$.

Note that $\Delta$ is the maximum of either the in- or out-degrees. Putting an upper bound on $\Delta$ rules out extremely large degrees, which are the trouble makers in the examples of Section 3.1. However, in most real networks, we have $\delta = 1$, so the sufficient condition essentially requires $\Delta < \sqrt{n}$, which, unfortunately, usually fails to hold. In fact, it fails for most networks studied in Section 5.

## 4. PATTERNS IN MAXIMUM DIGRAPHS

In this section, we identify some structural patterns of maximum digraphs, or equivalently, the associated suboptimal structures that contribute to the loss in reciprocity *not* imposed by the degree bi-sequence. We first look at some small suboptimal motifs and provide a greedy algorithm to eliminate them. We then show some more complicated structural patterns of maximum digraphs and demonstrate how they can help us pin down the maximum digraphs in some special cases. Omitted proofs can be found in [11].

Throughout this section, a cycle or a path always refers to a directed cycle or directed path, i.e. the edges must be all in the same direction as we follow the cycle or path. We also require that the edges be distinct. On the other hand, the vertices are not necessarily distinct. When the vertices are distinct, we say the path or cycle is elementary.

### 4.1 Small Suboptimal Motifs

In this subsection, we focus on a particular type of small motifs that we call 3-paths, the nonexistence of which also guarantees the nonexistence of many larger scale suboptimal structures. As we will see in Section 5, elimination of such suboptimal motifs brings reciprocity close to the corresponding upper bound for a variety of real world networks.

Given a digraph $G$, we call an elementary path of length 3, $\pi = (v_0, v_1, v_2, v_3)$, a 3-path if $(v_i, v_{i+1}) \in G_a$ for $i = 0, 1, 2$, i.e., $\pi$ consists entirely of unreciprocated edges. We further classify 3-paths into the following four types according to the connectivity between $v_0$ and $v_3$ (Figure 5),

(I). $(v_0, v_3) \notin G_u$, i.e. there is no edge between $v_0$ and $v_3$;

(II). $(v_0, v_3) \in G_s$;

(III). $(v_3, v_0) \in G_a$, i.e. $(v_0, v_1, v_2, v_3, v_0)$ is a 4-cycle;

(IV). $(v_0, v_3) \in G_a$.

As shown in Figure 5, 3-paths of Types I, II and III are suboptimal and can be rewired locally to increase reciprocity. We say a digraph is *3-path optimal* if it has no 3-path of Type I, II or III. Note that when viewed as a transformation on $G_a$, the rewiring procedure in Figure 5 simply eliminates 4-cycles (Type III), and replaces open 3-paths by a shortcut from its first vertex to its last vertex if such a shortcut does not yet exist (Types I and II). Thus each rewiring increases the number of reciprocated edges by either 2 or 4, and we have the following

LEMMA 1. *A maximum digraph is 3-path optimal.*

Given a digraph $G$, we can greedily rewire all 3-paths to get a lower bound on the maximum reciprocity allowed by the degree bi-sequence of $G$. The resulting greedy algorithm is shown in Algorithm 1. Lemma 2 guarantees that Algorithm 1 eliminates all 3-paths of Types I, II and III.

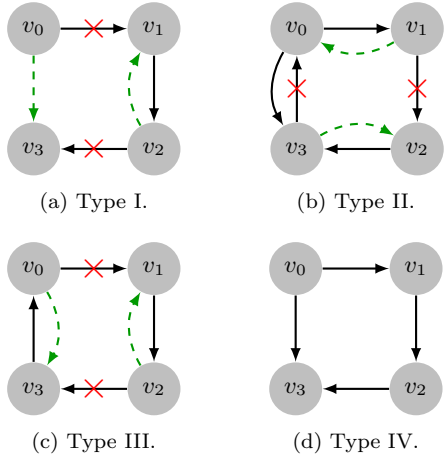

(a) Type I.  (b) Type II.

(c) Type III.  (d) Type IV.

Figure 5: Different types of 3-paths with corresponding rewirings. The edges marked by red crosses are to be rewired into the dashed green edges.

---

**Algorithm 1** GreedyRewire

**Input:** $G = (V, E)$
1: $S \leftarrow V$
2: **while** $S \neq \emptyset$ **do**
3:     pick $v_0 \in S$
4:     **if** $\exists$ non-Type IV 3-path $\pi = (v_0, v_1, v_2, v_3)$ **then**
5:        $G \leftarrow \text{Rewire}(\pi)$
6:        $S \leftarrow S \cup \{v_1, v_2\}$
7:     **else**
8:        $S \leftarrow S - \{v_0\}$
9:     **end if**
10: **end while**
11: **return** $G$

---

LEMMA 2. *Algorithm 1 returns a 3-path optimal digraph.*

Note that depending on how $v_0$ and $\pi$ are picked, Algorithm 1 can return different 3-path optimal graphs. Although there is no theoretical guarantee, we will see in Section 5 that reciprocities of 3-path optimal digraphs returned by Algorithm 1 are very close to the corresponding upper bounds and hence close to the maxima as well. The next subsection shows that 3-path optimality precludes many other suboptimal structures, which partially explains why Algorithm 1 works pretty well in practice.

## 4.2 Properties of Maximum Digraphs

In this subsection, we consider additional suboptimal structures that are more complicated than 3-paths. Some of these structures are automatically eliminated by Algorithm 1, while others require extra attention. We will state the results as properties of maximum digraphs. Any violation of the stated properties yields a suboptimal structure.

### 4.2.1 3-path optimal digraphs

We first consider some properties of 3-path optimal digraphs, which, by Lemma 1, are also properties of maximum digraphs. All these properties involve only unreciprocated edges. Note that any suboptimal structures that violate these properties are automatically eliminated by Algorithm 1. Let $G$ denote a 3-path optimal digraph throughout this subsection.

Lemma 3 shows that the unreciprocated edges of a 3-path optimal digraph cannot form any elementary path of odd length without a shortcut. As a result, for any two vertices $u$ and $v$, either there is no path from $u$ to $v$ in $G_a$, or there is such a path of length at most 2.

LEMMA 3. *If $\pi = (v_0, v_1, \ldots, v_{2p+1})$ is an elementary path of odd length in $G_a$, then $(v_0, v_{2p+1}) \in G_a$.*

Lemma 4 shows that the anti-symmetric subgraph of a 3-path optimal digraph is almost cycle free. We can obtain a directed acyclic graph from it by removing an edge from each 3-cycle.

LEMMA 4. *The only possible cycles in $G_a$ are 3-cycles, and any two of them must be vertex disjoint.*

Although 3-path optimality does not preclude 3-cycles, they are unlikely to exist in 3-path optimal graphs obtained from real world networks using Algorithm 1, as Lemma 5 requires that the vertices of a 3-cycle in such graphs have exactly the same connectivity to every vertex outside the 3-cycle, which is extremely unlikely, especially in large graphs.

LEMMA 5. *For a 3-cycle $C$ in $G_a$ and any vertex $v$ not in $C$, either there is no path in $G_a$ that connects $v$ and $C$, or there is an edge of $G_a$ between $v$ and each vertex of $C$, all in the same direction.*

### 4.2.2 Maximum digraphs

In this subsection, we consider some properties of maximum digraphs that are not direct consequences of 3-path optimality. The associated suboptimal structures may be left intact by Algorithm 1 and require extra attention. Throughout this subsection, let $G^\star$ denote a maximum digraph with a given bi-sequence $(\mathbf{d}^+, \mathbf{d}^-)$, i.e. $G^\star \in \mathcal{G}(\mathbf{d}^+, \mathbf{d}^-)$ and $\rho(G^\star) = \rho(\mathbf{d}^+, \mathbf{d}^-)$.

We know from Lemma 4 that large cycles involving only unreciprocated edges are suboptimal structures, but certain cycles of even length that contains reciprocated edges are also suboptimal. In particular, we have the following

LEMMA 6. *Let $C$ be an even cycle in $H \in \mathcal{G}(\mathbf{d}^+, \mathbf{d}^-)$. If any two edges in $C \cap H_s$ are separated by an odd number of edges in $C$, then there exists $H' \in \mathcal{G}(\mathbf{d}^+, \mathbf{d}^-)$ with $\rho(H') = \rho(H) + |C_a| - 2|C_a \cap H_s|$, where $C_a$ is the anti-symmetric part of $C$, i.e. $C_a = \{(i,j) \in C : (j,i) \notin C\}$.*

Note that $C \cap H_a \subset C_a$, but it is not necessarily true that $C_a = C \cap H_a$. The two edges $(3, 4)$ and $(5, 0)$ in Figure 6(a) are in $C_a$ but not in $C \cap H_a$. Any cycle satisfying the conditions in Lemma 6 is suboptimal if it has more anti-symmetric edges than symmetric ones. The cycles $(0, 1, 2, 3, 4, 5, 0)$ in Figure 6(a) and $(0, 1, 2, 0, 5, 3, 4, 5, 0)$ in Figure 6(b) are two such examples. Note that these two cycles are not automatically eliminated by Algorithm 1.

Lemma 7 specifies how multiple 3-cycles should be connected in maximum digraphs. If we collapse each 3-cycle into a single vertex by contracting its edges, the subgraph of $G^\star_a$ induced by these vertices will have the structure in Figure 4. Therefore, while the existence of multiple 3-cycles is already very unlikely in 3-path optimal digraphs, it is even less likely in maximum digraphs with degree bi-sequences of real world networks.

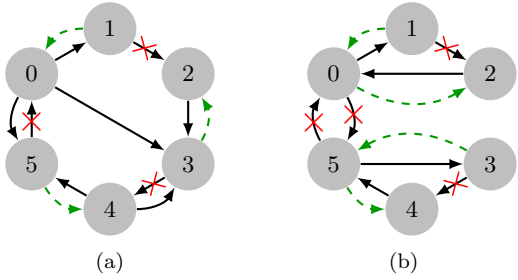

(a)  (b)

Figure 6: Suboptimal even cycle with reciprocated edges. Reciprocity can be increased by rewiring edges marked by red crosses into the dashed green edges.

LEMMA 7. *The set of all distinct 3-cycles of $G_a^\star$ can be linearly ordered as $C_0, C_1, \ldots, C_m$ such that there are 9 edges of $G_a^\star$ going from $C_i$ to $C_j$ for all $0 \le i < j \le m$.*

The next lemma complements Lemma 3 by specifying connection patterns of elementary paths of even length.

LEMMA 8. *Let $\pi = (v_0, v_1, \ldots, v_{2p})$ be an elementary path of even length $2p \ge 4$ in $G_a^\star$, $E_0 = \{(v_{2i}, v_{2j}) : i \ne j\}$ and $E_1 = \{(v_{2i-1}, v_{2j-1}) : i \ne j\}$. If $(v_0, v_{2p}) \notin G_a^\star$, then $G^\star$ either has all the edges in $E_0$ but none in $E_1$, or vice versa.*

Figure 7 shows both possibilities for an elementary path of length 4. The shortcuts required by Lemma 3 are also shown. The red dashed edges represent those that cannot coexist with the green edges in a maximum digraph. Some suboptimal structures that violate Lemma 8 cannot be automatically eliminated by Algorithm 1. For example, if the pair of edges between the vertices 0 and 2 are missing from Figure 7(a), the resulting suboptimal digraph will be left intact by Algorithm 1.

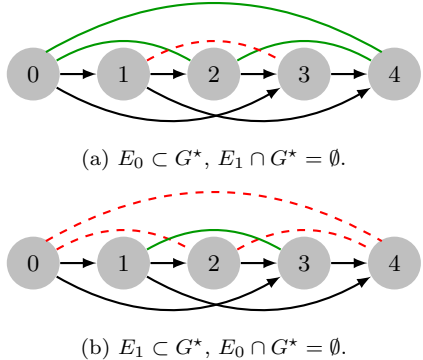

(a) $E_0 \subset G^\star$, $E_1 \cap G^\star = \emptyset$.

(b) $E_1 \subset G^\star$, $E_0 \cap G^\star = \emptyset$.

Figure 7: Patterns of even paths in maximum digraphs.. Each undirected solid edge represents a pair of reciprocated edges in $G^\star$. Each dashed edge represents a pair of edges that are both missing in $G^\star$.

## 4.3 Some Examples

In this subsection, we illustrate how the structural patterns of the previous subsection may be used to pin down the maximum digraph in some special cases. Here $G^\star$ always denotes a maximum digraph.

Proposition 2 shows that when the bi-sequence is perfectly balanced, the maximum digraph achieves perfect or near-perfect reciprocity. Therefore, any unfulfilled reciprocity must be due to the lack of effort to form reciprocal edges rather than due to the fundamental limit imposed by the bi-sequence itself.

PROPOSITION 2. *Suppose $(\mathbf{d}^+, \mathbf{d}^-)$ is perfectly balanced, i.e. $\nu(\mathbf{d}^+, \mathbf{d}^-) = 0$.*

(1). *If $\varepsilon(\mathbf{d}^+, \mathbf{d}^-)$ is even, then $\rho(\mathbf{d}^+, \mathbf{d}^-) = \varepsilon(\mathbf{d}^+, \mathbf{d}^-)$.*

(2). *If $\varepsilon(\mathbf{d}^+, \mathbf{d}^-)$ is odd, then $\rho(\mathbf{d}^+, \mathbf{d}^-) = \varepsilon(\mathbf{d}^+, \mathbf{d}^-) - 3$, and $G_a^\star$ consists of a 3-cycle.*

PROOF. Since $\nu(\mathbf{d}^+, \mathbf{d}^-) = 0$, we have $d_i^+ = d_i^-$ for all $i$. Thus any edge of $G_a^\star$ must be contained in a cycle of length at least 3 in $G_a^\star$. By Lemma 4, the length of such a cycle is exactly 3. By Lemma 7, there is at most one such cycle in $G_a^\star$. Thus $G_a^\star$ is either empty or a 3-cycle. Since $\rho(G^\star)$ must be even, the former case corresponds to even $\varepsilon(\mathbf{d}^+, \mathbf{d}^-)$ and the latter odd $\varepsilon(\mathbf{d}^+, \mathbf{d}^-)$. □

The next proposition shows that when the bi-sequence is slightly unbalanced, the number of possible values of $\rho(\mathbf{d}^+, \mathbf{d}^-)$ increases. This sheds some light on why the maximum reciprocity problem is so difficult. As the total unbalanced degree increases, the number of possibilities is expected to explode.

PROPOSITION 3. *Suppose $(\mathbf{d}^+, \mathbf{d}^-)$ is slightly unbalanced with $\nu(\mathbf{d}^+, \mathbf{d}^-) = 1$, $d_0^+ - d_0^- = 1$ and $d_1^- - d_1^+ = 1$.*

(1). *If $\varepsilon(\mathbf{d}^+, \mathbf{d}^-)$ is even, then the gap $\varepsilon(\mathbf{d}^+, \mathbf{d}^-) - \rho(\mathbf{d}^+, \mathbf{d}^-)$ is either 2 or 4. When the gap is 2, the two edges in $G_a^\star$ form a 2-path from 0 to 1. When the gap is 4, $G_a^\star$ is the vertex disjoint union of $\{(0,1)\}$ and a 3-cycle.*

(2). *If $\varepsilon(\mathbf{d}^+, \mathbf{d}^-)$ is odd, then the gap $\varepsilon(\mathbf{d}^+, \mathbf{d}^-) - \rho(\mathbf{d}^+, \mathbf{d}^-)$ is either 1 or 5. When the gap is 1, $G_a^\star = \{(0,1)\}$. When the gap is 5, $G_a^\star$ is the vertex disjoint union of a 2-path from 0 to 1 and a 3-cycle.*

PROOF. Note that there must be a path from 0 to 1 in $G_a^\star$. Let $\pi$ be the shortest path from 0 to 1 in $G_a^\star$. All edges in $G_a^\star - \pi$, if there is any, must be contained in a cycle in $G_a^\star$. By Lemma 4, $G_a^\star$ can only have 3-cycles. If $G_a^\star$ had more than one 3-cycles, Lemma 7 would require that there be at least 9 edges in $G_a^\star$ that are not contained in any cycle, all of which must be in $\pi$. Lemma 3 shows that $\pi$ has either one or two edges. Therefore, $G_a^\star - \pi$ is either empty or has one 3-cycle. By Lemma 5, $\pi$ and the 3-cycle, if there is one, must be vertex disjoint. Since $|\pi| \in \{1, 2\}$, and $|G_a^\star - \pi| \in \{0, 3\}$, it follows that $\varepsilon(\mathbf{d}^+, \mathbf{d}^-) - \rho(G^\star) = |G_a^\star| = |\pi| + |G_a^\star - \pi| \le 2 + 3 = 5$. Note that $\rho(G^\star)$ is even. If $\varepsilon(\mathbf{d}^+, \mathbf{d}^-)$ is even, then $\varepsilon(\mathbf{d}^+, \mathbf{d}^-) - \rho(G^\star)$ is equal to $|\pi| = 2$ or $\alpha(G^\star) = |\pi| + |G_a - \pi| = 1 + 3 = 4$. If $\varepsilon(\mathbf{d}^+, \mathbf{d}^-)$ is odd, then $\varepsilon(\mathbf{d}^+, \mathbf{d}^-) - \rho(G^\star)$ is equal to $|\pi| = 1$ or $|\pi| + |G_a^\star - \pi| = 2 + 3 = 5$. □

It is easy to come up with examples where the gaps are 1 and 2, respectively. The next examples shows that the other two cases are also possible.

*Example 5.* Let $(\mathbf{d}^+, \mathbf{d}^-) = \{(1, 3, 2, 2, 2), (0, 4, 2, 2, 2)\}$. Figure 8(a) shows a realization $G$ of this bi-sequence, where each undirected edge represents a pair of edges in opposite directions. Note that $\rho(G) = \varepsilon(\mathbf{d}^+, \mathbf{d}^-) - 4$. We claim that $\rho(G) = \rho(\mathbf{d}^+, \mathbf{d}^-)$. If not, then $\rho(G^\star) = \varepsilon(\mathbf{d}^+, \mathbf{d}^-) - 2$ by Proposition 3, and the two edges in $G_a^\star$ form a 2-path $\pi$ from $a$ to $b$. Since $c$, $d$, $e$ have the same in- and out-degrees and hence are equivalent, we may assume without loss of generality that $\pi = (a, c, b)$. Thus $G_a^\star - \pi$ is symmetric and corresponds to a simple graph with degree sequence $\hat{\mathbf{d}} = \{0, 3, 1, 2, 2\}$. There is only one simple graph with this degree sequence, which is shown by the black edges in Figure 8(b). When we superimpose $\pi$ and $G_a^\star - \pi$, there are two edges from $(c, b)$, and hence $G^\star \notin \mathcal{G}(\mathbf{d}^+, \mathbf{d}^-)$, a contradiction.

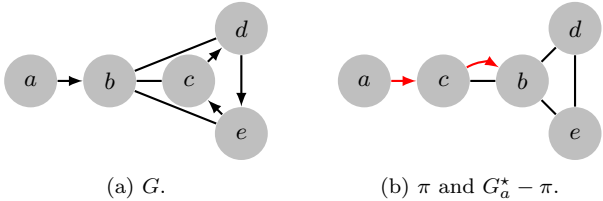

Figure 8: Example 5.

*Example 6.* Let $(\mathbf{d}^+, \mathbf{d}^-) = \{(1, 0, 4, 2, 2, 2), (0, 1, 4, 2, 2, 2)\}$. Figure 9 shows a realization $G$ of this bi-sequence, where each undirected edge represents a pair of edges in opposite directions. Note that $\rho(G) = \varepsilon(\mathbf{d}^+, \mathbf{d}^-) - 5$. Since the sequence $\mathbf{d}^+ \wedge \mathbf{d}^- = \{0, 0, 4, 2, 2, 2\}$ is not graphic, Proposition 1 shows that $\rho(\mathbf{d}^+, \mathbf{d}^-) < \beta(\mathbf{d}^+, \mathbf{d}^-) = \varepsilon(\mathbf{d}^+, \mathbf{d}^-) - 1$. Thus Proposition 3 yields $\rho(G) = \rho(\mathbf{d}^+, \mathbf{d}^-)$. In fact, $G$ is the only element of $\mathcal{G}(\mathbf{d}^+, \mathbf{d}^-)$.

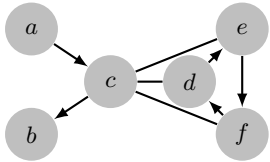

Figure 9: Example 6.

## 5. EMPIRICAL STUDY

In this section, we conduct an empirical analysis of real networks by comparing the observed values of reciprocity against the upper bounds. We also look at the lower bounds on maximum reciprocities given by Algorithm 1.

### 5.1 Datasets

The networks that we analyze include major online social networks (OSN) that are directed in nature [15, 12, 8, 24, 13]. For the purpose of comparison, we have also included other types of networks: biological networks [20, 21, 18, 23, 19, 27], communication networks [13], product co-purchasing networks [13], web graphs [13], Wikipedias [1], software call graphs [22, 17], and P2P networks [13]. All the datasets except for Wikipedias are already converted into graph representations by other researchers and the descriptions for the datasets can be found at the cited sources. For Wikipedias, each node represents a page. Only article pages, i.e. pages with namespace ID 0, are included. Pages that redirect to the same page are represented as a single node corresponding to the destination page. There is an edge from node $A$ to node $B$ if there is at least one hyperlink from page $A$ to page $B$. Multiple edges and self-loops are discarded.

### 5.2 Empirical Reciprocity vs. Upper Bound

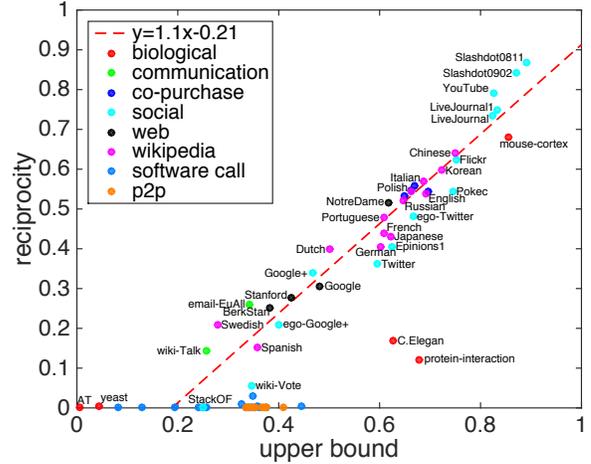

Figure 10: Scatter plot of empirical reciprocity versus upper bound. Regression line was fitted without data points for biological, P2P and software call networks.

Figure 10 shows the scatter plot of empirical reciprocities against the corresponding upper bounds. Here the upper bound is normalized by the number of edges, i.e., it is the ratio $\beta(\mathbf{d}^+, \mathbf{d}^-)/\varepsilon(\mathbf{d}^+, \mathbf{d}^-)$. Note that the reciprocity values vary widely, ranging from 0 for the peer-to-peer network Gnutella to 90% for the online social network Slashdot. There is even a fair amount of variation within the categories of biological, social and Wikipedia networks. In general, social networks and Wikipedia networks tend to have high reciprocity, while software call networks tend to have low reciprocity. Note the strong linear correlation between empirical reciprocity and the upper bound. This is a bit surprising, especially for the social networks, in view of the large variations in reciprocity. Related to Figure 10 is the scatter plot in Figure 11 of number of reciprocated edges against the unnormalized bound $\beta(\mathbf{d}^+, \mathbf{d}^-)$. There the linear relationship in log-log scale is even more apparent, with biological networks being also around the regression line. These linear relationships suggest that there might exist some universal mechanism that works across different domains.

Despite the wide variation in reciprocity, the ratio between the empirical reciprocity and the normalized upper bound has a much narrower range as shown by the box plots for the ratios in Figure 12.

Note that the ratios are close to zero for the P2P network Gnutella and software call graphs. The Gnutella exhibits zero reciprocity, far away from the upper bounds, which are above 30%. This is probably because Gnutella implements an indirect reciprocity mechanism. The low reciprocity for software call graphs is not surprising, as software codes are usually designed to work in a hierarchical manner. The case for biological networks are more complicated, as the four biological networks considered here are actually of quite different natures. For example, the C. Elegan neural network and the mouse cortex network are both neural networks, but the former is at the neuron level while the latter is at a coarser level of cortical regions. One can speculate that both the low reciprocity in C. Elegan neural network and the high reciprocity in the mouse cortex network are due to biological reasons. However, we do not know if this behavior is a norm or an exception due to the lack of data for similar networks.

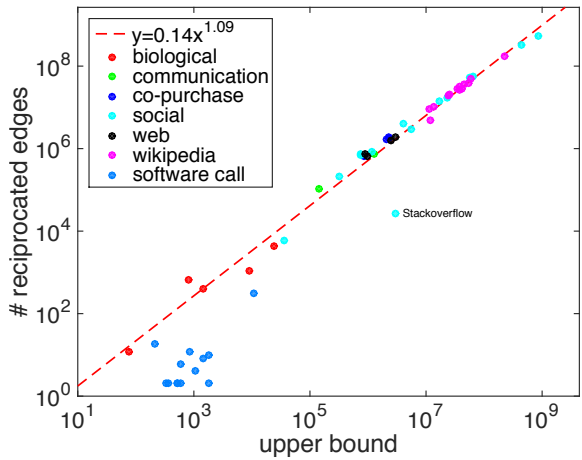

Figure 11: Scatter plot of number of reciprocated edges versus upper bound. Regression line was fitted in log scale, without data points for software call networks.

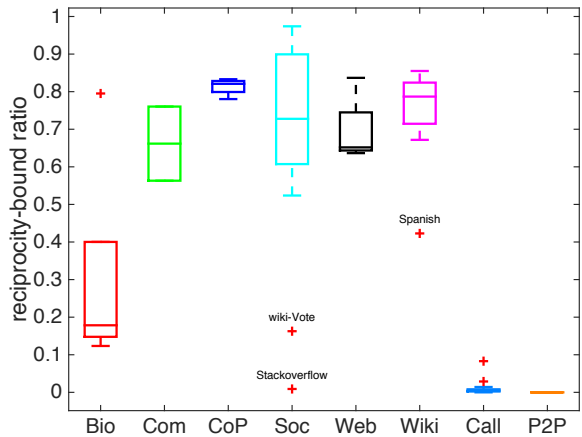

Figure 12: Box plot of reciprocity-bound ratio for different network categories.

In all categories other than biological, software call and P2P networks, the ratios are above 50% with only three exceptions: the wiki-Vote network, the Stack Overflow Q&A network, and the Spanish Wikipedia. Although we have classified the Stack Overflow Q&A network as a social network, it differs from typical social networks. The low reciprocity suggests that there is a hierarchy of expertise. What is more interesting is the wiki-Vote network and the Spanish Wikipedia, as their behaviors deviate from those of other networks of the same category, which suggests that there might be something unusual about them that is worthy of scientific study. Apart from the three outliers, all other networks in these categories actually achieve a significant fraction of the possible reciprocity suggested by the upper bound. This means that modulo the degree constraints, the tendency to reciprocate is much stronger than the empirical reciprocity alone might have suggested. Prominent examples include the web graphs, the Swedish Wikipedia and the Google+ network, whose reciprocities are not very high in absolute value but quite high relative to the bound. This suggests that when we study these networks, it might be more meaningful to ask the question why there is such large imbalance in degrees than to ask the question why the tendency to reciprocate is low.

## 5.3 Reciprocity of 3-path Optimal Digraphs

In this subsection, we look at 3-path optimal digraphs returned by Algorithm 1. Note that the reciprocity of such a digraph provides a lower bound on the maximum reciprocity of the corresponding degree bi-sequence.

Figure 13 shows the scatter plot of the reciprocities of the 3-path optimal digraphs against the corresponding upper bounds. Note that the reciprocities of 3-path optimal digraphs are very close to the upper bounds, which means that the maximum reciprocities are also very close to the upper bounds. Therefore, for the degree bi-sequences of those real networks, the fundamental limit that they impose on reciprocity is largely summarized by the upper bounds, and the major source of loss in reciprocity is the existence of 3-paths of Types I, II and III. Thus in practice Algorithm 1 usually suffices for approximating maximum reciprocities and we do not need to worry much about the more complicated suboptimal structures in Section 4.2.2.

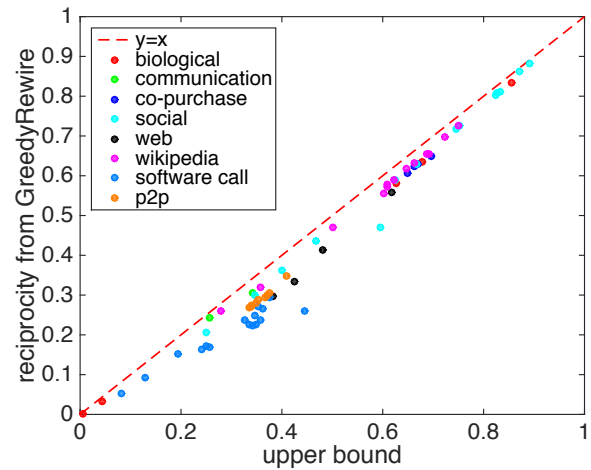

Figure 13: Scatter plot of reciprocity of the 3-path optimal digraph returned by Algorithm 1 versus upper bound.

Finally, recall from Section 4.2.1 that the existence of 3-cycles in a 3-path optimal digraph requires some specific structures. These structures are usually too special to occur in practice, so 3-cycles are unlikely to exist in 3-path optimal digraphs. This is indeed the case for most of the 3-path optimal digraphs obtained from the real networks studied here, the anti-symmetric parts of which turn out to be acyclic.

## 6. CONCLUSION

In this work, we showed that the maximum reciprocity problem is NP-hard. We provided a partial characterization of networks with maximum reciprocity and a greedy algorithm to eliminate suboptimal motifs. We also provided an upper bound on reciprocity along with necessary conditions and sufficient conditions for achieving the bound. We demonstrated that the bound is surprisingly close to the observed reciprocity in a wide range of real networks, which suggests that the tendency to form reciprocal edges might be much stronger than the observed reciprocity indicates. We found surprising linear relationships between empirical reciprocities and the corresponding upper bounds. We showed that a particular type of suboptimal motif called 3-paths is the major source of loss in reciprocity in these networks.

## Acknowledgements

This work was supported in part by DoD ARO MURI Award W911NF-12-1-0385, NSF grants CNS-1065133, CNS-1117536, CNS-1411636, and DTRA grants HDTRA1-09-1-0050 and HDTRA1-14-1-0040.